\documentclass[twocolumn,3p]{elsarticle}
\usepackage{colordvi,epsfig,color,amsmath,amssymb}

\begin{document}

\def\be{\begin{equation}}
\def\ee{\end{equation}}
\def\bee{\begin{eqnarray}}
\def\eee{\end{eqnarray}}
\def\sech{\mbox{sech}}
\def\e{{\rm e}}
\def\d{{\rm d}}
\def\L{{\cal L}}
\def\U{{\cal U}}
\def\M{{\cal M}}
\def\T{{\cal T}}
\def\V{{\cal V}}
\def\R{{\cal R}}
\def\kb{k_{\rm B}}
\def\tw{t_{\rm w}}
\def\ts{t_{\rm s}}
\def\Tc{T_{\rm c}}
\def\gs{\gamma_{\rm s}}
\def\tm{tunneling model }
\def\TM{tunneling model }
\def\tilde{\widetilde}
\def\Deltac{\Delta_{0\rm c}}
\def\Deltamin{\Delta_{0\rm min}}
\def\Emin{E_{\rm min}}
\def\tauc{\tau_{\rm c}}
\def\tauac{\tau_{\rm AC}}
\def\tauw{\tau_{\rm w}}
\def\taumin{\tau_{\rm min}}
\def\taumax{\tau_{\rm max}}
\def\de{\delta\varepsilon / \varepsilon}
\def\pF{{\bf pF}}
\def\pFAC{{\bf pF}_{\rm AC}}
\def\halb{\mbox{$\frac{1}{2}$}}
\def\with{\quad\mbox{with}\quad}
\def\und{\quad\mbox{and}\quad}
\def\za{\sigma_z^{(1)}}
\def\zb{\sigma_z^{(2)}}
\def\ya{\sigma_y^{(1)}}
\def\yb{\sigma_y^{(2)}}
\def\xa{\sigma_x^{(1)}}
\def\xb{\sigma_x^{(2)}}
\def\spur#1{\mbox{Tr}\left\{ #1\right\}}
\def\erwart#1{\left\langle #1 \right\rangle}
\newcommand{\bbbone}{{\mathchoice {\rm 1\mskip -4mu l}{\rm 1\mskip -4mu l}{\rm
1\mskip -4.5mu l}{\rm 1\mskip -5mu l}}}

\title{Competition between relaxation and external driving in the dissipative
Landau-Zener problem \tnoteref{peter}}

\tnotetext[peter]{Dedicated to the 60th birthday of Peter H\"anggi}

\author[ham]{P.~Nalbach\corref{cor1}}
\ead{Peter.Nalbach@physik.uni-hamburg.de}
\author[ham]{M.~Thorwart}
\ead{Michael.Thorwart@physik.uni-hamburg.de}
\cortext[cor1]{Corresponding author}
\address[ham]{I.\ Institut f\"ur Theoretische Physik,  Universit\"at Hamburg,
Jungiusstra{\ss}e 9, 20355 Hamburg, Germany}

\begin{abstract}

We study Landau-Zener transitions in a dissipative environment by means of the
 quasiadiabatic propagator path-integral scheme.
It allows to obtain numerically exact results for the full range of the involved parameters.
We discover a nonmonotonic dependence of the Landau-Zener transition probability on the sweep velocity which is explained in terms of a simple physical picture. This feature results from a nontrivial competition between relaxation processes and the external sweep and is not captured by perturbative approaches. In addition to the Landau-Zener transition probability, we study the excitation survival probability and also provide a qualitative understanding of the involved competition of time scales.

\end{abstract}


\maketitle

\section{Introduction}

The transition dynamics in a quantum two-level system with a time-dependent
Hamiltonian
varying such that the energy separation of the two diabatic states is a linear function
of time is
a central problem since the early days of quantum mechanics. It is commonly
denoted as the Landau-Zener (LZ) problem, although it has been solved
independently
by Landau \cite{LZLa1932}, Zener \cite{LZZe1932}, St\"uckelberg \cite{LZSt1932}
and Majorana \cite{LZMa1932} in 1932 (for a more detailed discussion of the
differences between the
four approaches, we refer the interested reader to a paper by Di Giacomo and
Nikitin \cite{Giacomo}).
Nonadiabatic transitions at avoided level crossings are at the heart of
many dynamical processes throughout physics and chemistry. They have been
extensively studied both theoretically and experimentally in, e.g., spin flips
in nanomagnets \cite{Wernsdorfer}, solid state artificial atoms
\cite{LZSi2006,LZBe2008,LZKi2008}, nanocircuit QED \cite{QED1,QED2},
adiabatic quantum computation \cite{AQC}, the dynamics of chemical
reactions \cite{Nitzan}, and in Bose-Einstein condensates in optical
lattices \cite{Zenesini2009}. Also the nonequilibrium dynamics of glasses
at low temperatures is dominated by the transition dynamics of avoided level
crossings~\cite{TSGLNEQRo2003,TSGLNEQLu2003,TSGLNEQNa2004,TSGLNEQNa2005}.

In the pure Landau-Zener problem, two quantum states interact by a constant
tunneling
matrix element $\Delta_0$. A control parameter is swept through the avoided
level crossing at a constant velocity $v$, such that the
energy gap between the two diabatic states depends linearly on time.
The Landau-Zener problem addresses the
case when the system starts in the lower energy eigenstate in the infinite past
and asks for the probability of
finding the system in the lower energy eigenstate in the infinite future
(a Landau-Zener transition). Certainly, for infinitely slow variation of the
energy difference $v \to 0$, the adiabatic theorem states that no transition
between energy eigenstates will occur, since at any moment of time,
the system will always be in an instantaneous eigenstate of
the Hamiltonian.
For $v\ne 0$, the probability $P_0$ for no transition~\cite{foot1} is 
described by the Landau-Zener formula \cite{LZLa1932,LZZe1932,LZSt1932,LZMa1932}
$P_0(v,\Delta_0) = 1 - \exp[ -\pi\Delta_0^2/(2v)]$.

Most experimental investigations, on the one hand, actually deal with
more complex systems than two-state systems and, on the other
hand, do not necessarily start in the ground state.
In close proximity around the avoided crossing in most cases only
the two crossing states are important to describe the dynamics.
However, in general in such an approximate system we also have to deal
with the system being initially in the excited state.
In the pure Landau-Zener problem the probability to end up in the
ground (excited) state when starting in the ground (excited) state
are identical due to symmetry.

However, in any physical realization, a quantum system is influenced by
its environment leading to relaxation and phase decoherence during time
evolution \cite{WeissBuch,SpiBoLe1987}.
At low temperatures when the environmental fluctuations are not thermally
occupied, the Landau-Zener probability $P$ (to end up in the ground state
when starting in the ground state) should hardly be influenced since no
phonons are available for the system to be in the excited state at 
asymptotically long times. However,
spontaneous emission is possible even at lowest temperatures and the
excitation survival probability $Q$ (to end up in the excited state
when starting in the excited state) is expected to decrease when the
system has the time to relax during the driving. Thus especially at
low driving speeds, the excitation survival probability will be reduced
since the system will decay, while the pure Landau-Zener mechanism predicts
full survival of excitations when driving through an avoided crossing
at low speeds.


The dissipative Landau-Zener problem received a lot of
attention~\cite{LZKaI1984,LZKaII1984,LZAo1991,LZWu2006,LZSa2007,LZKa1998,
LZPo2007,LZNa2009} in the past 25 years due to its relevance for controlled
quantum state preparation which became experimentally feasible in many physical
realizations. Although the full problem is analytically unsolved, many limiting
cases are analytically tractable and numerical approaches are available. Usually
a dissipative environment causes fluctuations of the energies of the diabatic
states (longitudinal or diagonal system-bath coupling) but occasionally the
environments can also cause transitions between the diabatic states (transversal
system bath coupling). In this work, we focus on the more common case of longitudinal 
coupling. The transversal coupling has been treated in Refs.\ 
\cite{LZAo1991,LZWu2006} for zero temperature. 

In the limit of very fast sweeps (nonadiabatic driving), a thermal heat bath has
been
shown not to influence the Landau-Zener probability. This follows consistently
from all approaches (for the corresponding limiting conditions of the other free
parameters)~\cite{LZKaI1984,LZKaII1984,LZAo1991,LZWu2006,LZSa2007,LZKa1998,
LZPo2007,LZNa2009}.

At low temperatures, the Landau-Zener (or, equivalently, ground state survival)
probability $P$
(to end up in the ground state when starting in the ground state) is only
influenced little
by a diagonally coupled bath. This was first shown by Ao and
Rammer~\cite{LZAo1991}.
Kayanuma~\cite{LZKaI1984,LZKaII1984} showed before that in the limit of slow
fluctuations
of the diabatic energies (as might be caused by a diagonally coupled bath at
low temperatures),
the Landau-Zener probability is not modified. At zero temperature, a diagonally
coupled bath
has strictly no influence on the Landau-Zener probability, as has been shown by
Wubs {\em et al.\/}~\cite{LZWu2006}. This is true not only for bosonic but also for spin
environments~\cite{LZSa2007}.
In contrast, transversely coupled baths lower the Landau-Zener probability
even at zero temperatures~\cite{LZWu2006} and their effects depend on the
details of the bath characteristics~\cite{LZSa2007}.

The strong damping limit can be reached in two ways. For large coupling
between system and bath and finite temperatures, Ao and Rammer~\cite{LZAo1991}
have shown that again a diagonally coupled bath does not influence the Landau-Zener
probability. Conversely, in the high temperature limit under adiabatic conditions
(i.e., slow sweeps), the two states are driven to equal population, $P=P_{\rm
SD}=\halb(1-\exp(\pi\Delta_0^2/v))$. This has first been shown by
Kayanuma~\cite{LZKaI1984,LZKaII1984} for large and fast fluctuations and later
by Ao and Rammer~\cite{LZAo1991}. Pokrovsky and Sun~\cite{LZPo2007} have extended
Kayanuma's result to transversely coupled environments.

For the experimentally important parameter range of slow (adiabatic) sweeps and
intermediate temperatures but only weakly coupled environments, Pokrovsky and
Sun~\cite{LZPo2007}, Kayanuma and Nakayama~\cite{LZKa1998} and Ao and
Rammer~\cite{LZAo1991} each give approximate solutions of the influence of the
environmental fluctuations on the Landau-Zener probability. Using the numerically exact
quasiadiabatic propagator path-integral (QUAPI) technique,
we recently described the Landau-Zener probability in
the full parameter space~\cite{LZNa2009}. For small sweep velocities and medium
to high temperatures, we have discovered non-monotonic dependencies on the sweep
velocity, temperature, coupling strength and cut-off frequency which were not
included in the previous approximate solutions. This behavior can be understood
in simple physical terms as a nontrivial competition between relaxation and
Landau-Zener driving.

The direct influence of environmental fluctuations on the dynamics of a driven
two-state system is much more pronounced when the system is initially prepared in the
excited state since spontaneous emission is even possible at zero temperature.
Accordingly,
the excitation survival probability $Q$ (to end up in the excited state having started
in the excited state), which is without environment strictly identical
to the Landau-Zener probability, is strongly modified for all temperatures and
bath coupling strengths~\cite{LZAo1991,LZKa1998}. Although experimentally
highly relevant, the excitation survival probability received much less
attention. In the limit of fast (nonadiabatic) sweeps, no influence of a bath is
expected~\cite{LZAo1991,LZKa1998}. In the high temperature limit under adiabatic
conditions (slow sweeps), the two states are again driven to equal population,
$Q_{\rm SD}=P_{\rm SD}=\halb(1-\exp(\pi\Delta_0^2/v))$~\cite{LZAo1991,LZKa1998}.
Thus, in both limits the excitation survival probability is identical to
the Landau-Zener probability, $Q=P$. However, this is quite different at
intermediate and low temperatures since spontaneous emission drastically changes
the excitation survival probability. Ao and Rammer~\cite{LZAo1991} were the
first to remark that in the limit of weak system-bath coupling, a
discontinuity occurs. In the limit of strong coupling and low
temperature, Kayanuma and Nakayama~\cite{LZKa1998} find another simple
analytical expression, $Q_{\rm sc} =P_0(1-P_0)$ with $P_0$ being the Landau-Zener
probability without environment. Both, Kayanuma and Nakayama~\cite{LZKa1998}
and Ao and Rammer~\cite{LZAo1991} give approximate solutions to the experimentally
important parameter range of slow (adiabatic) sweeps and intermediate
temperatures, but only weakly coupled environments. Nevertheless,
this parameter range is still largely unexplored.

Beyond these direct approaches, the dissipative Landau-Zener problem was
discussed in many more facets. Not being able to give a full review of all
works, we just mention in passing that Moyer
discussed the Landau-Zener problem for decaying states involving complex energy {\it
eigenvalues\/}~\cite{LZMo2001}. Extensions to three-state systems~\cite{LZSa2002}
or circuit QED problems~\cite{LZZu2008} have been made recently.
Finally, we also mention that spin environments in context of Landau-Zener transitions
have been discussed by Garanin et al.~\cite{LZGa2008}.


In this paper, we investigate the dissipative Landau-Zener problem in the full
parameter range of sweep velocities, temperatures, damping strengths and cut-off
frequency by means of the quasiadiabatic propagator path-integral (QUAPI)
\cite{QUAPI1,QUAPI2,QUAPI3,QUAPI4,QUAPI5}.
It allows to include nonadiabatic as well as non-Markovian effects yielding
numerically exact results.
In the next section we introduce the basic model. In the third section we
discuss the time-dependent occupation probability of the two states during a
Landau-Zener transition and in the following section we focus on the asymptotic populations
discussing the Landau-Zener (ground state survival) and the excitation survival
probabilities in dependence of the sweep velocities, temperatures, damping strengths
and cut-off frequency. We show that interesting features arise due to a competition
between time scales associated to the Landau-Zener sweep and to dissipative transitions.
Finally, we conclude with a short summary.

\section{Model}

\begin{figure}[t]
\epsfig{file=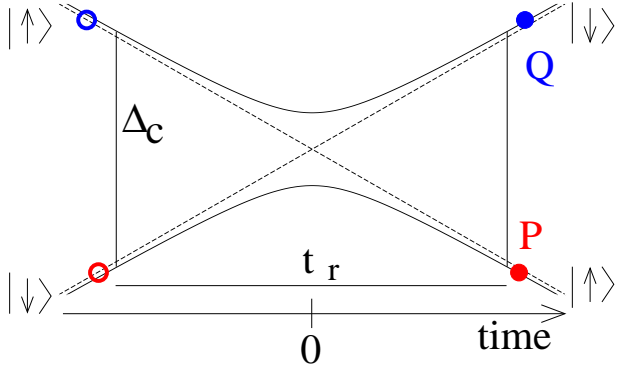,width=8cm}
\caption{\label{figH1} The spectrum of a two-level system driven linearly in
time through an avoided crossing. The solid lines mark
the exact energies whereas the dashed lines give the crossing energies $\pm vt$ of the
diabatic states. In addition, the resonance time window of width $t_r$ around $t=0$ is shown, in
which $\Delta_t\le\Delta_c$, see Sec.~\ref{physpic} for more details.}
\end{figure}

A quantum mechanical two-state system which shows an avoided energy
level crossing when driven is described
by the Landau-Zener Hamiltonian ($\hbar=1$)
\be
H_{LZ}(t) = \frac{\Delta_0}{2}\sigma_{x} +\frac{vt}{2} \sigma_z \, ,
\ee
with the tunneling matrix element $\Delta_0$ and the energy gap between the
diabatic states $vt$, changing linearly in time with sweep velocity $v$.
Here, $\sigma_{x,z}$ are Pauli matrices and the diabatic states are the
eigenstates
($|\downarrow\rangle$ and $|\uparrow\rangle$)
of $\sigma_z$. Asymptotically at times $|t|\gg\Delta_0/v$,
the diabatic states coincide with the momentary eigenstates of $H_{LZ}$.
Figure~\ref{figH1} plots the eigenenergies of $H_{LZ}$ (full lines) which show
an avoided level crossing with minimal splitting $\Delta_0$ and the energies of
the diabatic states (dashed lines) which form an exact crossing as a function of
 time.

The Landau-Zener problem asks for the probability $P_0$ of the system to end up
in the ground state at $t=+\infty$, having started in the ground state at
$t=-\infty$ (the corresponding one to end in the excited state is given as
$1-P_0$). Its exact solution dates back to the year 1932
\cite{LZLa1932,LZZe1932,LZSt1932,LZMa1932} and is given by
\be
\hspace*{-1mm}P_0(v,
\Delta_0)=|\langle\uparrow(\infty)|\downarrow(-\infty)\rangle|^2 = 1 -
\exp\left( -\frac{\pi\Delta_0^2}{2v} \right)
\ee

The excitation survival probability $Q_0$ to end up in the excited state at
$t=+\infty$, having started in the excited state at $t=-\infty$ is strictly
identical, $Q_0=P_0$ for the coherent two-state problem.

To include environmental fluctuations on Landau-Zener transitions, we couple
$H_{LZ}$ diagonally to a harmonic bath \cite{WeissBuch,SpiBoLe1987}, yielding
\be H(t) = H_{LZ}(t) -\frac{\sigma_z}{2}\sum_k\lambda_k (b_k+b_k^\dagger)
+\sum_k \omega_k \left(b_k^\dagger b_k+\halb\right)
\ee
with the bosonic annihilation/creation operators $b_k/b_k^\dagger$.
The bath influence is captured by the spectral function $J(\omega)=2\alpha\omega
\exp(-\omega/\omega_c)$, for which
we choose here for definiteness an Ohmic form
with the cut-off frequency $\omega_c$ and the coupling strength $\alpha$
\cite{WeissBuch,SpiBoLe1987}.
The Landau-Zener probability for the dissipative problem
$P=\spur{|\uparrow\rangle\langle\uparrow|U_\infty|\downarrow\rangle\langle\downarrow|U^{-1}
_\infty}$
with the time evolution operator $U_\infty={\cal T}
\exp[-i\int_{-\infty}^\infty dt H(t)]$ as well as the excitation survival
probability
$Q=\spur{|\downarrow\rangle\langle\downarrow|U_\infty|\uparrow\rangle\langle\uparrow|U^{-1}
_\infty}$
are now functions not only of $\Delta_0$ and $v$, but also of $\alpha, \omega_c$
and the temperature $T$.
In the following, we use $\omega_c=10\Delta_0$ unless specified otherwise.

\section{Time dependent occupation probabilities}

In this section, we explicitly consider the time-dependence of the
population of the diabatic state $|\uparrow\rangle$ at any instant of time $t$, having
started in $|\downarrow \rangle$ with probability one. This
is given by
\be
P(t)=\spur{|\uparrow\rangle\langle\uparrow|U_t|\downarrow\rangle\langle\downarrow|U^{-1}_t}
\ee
with the time evolution operator $U_t={\cal T}
\exp[-i\int_{-\infty}^t dt' H(t')]$. Note that at asymptotic times, this quantity
coincides with the standard Landau-Zener probability $P=P(t\to \infty)$.
Note furthermore that $P(t)$ is the tunneling probability at any instant of time.
Here, the dynamics of the quantum two-level system is described
in terms of the time evolution of the reduced density matrix $\rho(t)={\rm Tr}_{B}
\{U_t \rho_0 U^{-1}_t\}$, starting
from a total initial density matrix $\rho_0=\rho_S \otimes e^{-H_B/T}/Z_B$, where
$\rho_S$ is the density operator of the quantum two-level system and $H_B$ denotes the
bath Hamiltonian which is assumed to be decoupled at $t_0=-\infty$ (which is in practice set to zero)
and instantly switched on directly afterwards. Moreover, $Z_B=\spur{e^{-H_B/T}}$ with $k_B=1$.
$\rho(t)$ is obtained after
tracing over the bath degrees of freedom. We calculate $\rho (t)$ using the numerically
exact quasiadiabatic propagator path-integral \cite{QUAPI1,QUAPI2,QUAPI3,QUAPI4,QUAPI5} scheme.
For details of the iterative technique, we refer to previous works \cite{QUAPI1,QUAPI2,QUAPI3,QUAPI4,QUAPI5}.

In brief, the algorithm is based on a symmetric Trotter splitting
of the short-time propagator ${\cal K}(t_{k + 1}, t_k)$ for the full Hamiltonian
into a part depending on the system Hamiltonian and a part involving the
bath and the coupling parts. The short time propagator describes time evolution
over a Trotter time slice $\Delta t$. This splitting is of course exact in the limit
$\Delta t \to 0$ but introduces a finite Trotter error to the splitting,
which has to be eliminated by choosing $\Delta t$ small enough such that convergence
is achieved. On the other hand, the bath degrees of freedom generates correlations
being non-local in time. For any finite temperature, these correlations
decay exponentially fast at asymptotic times, thereby defining the associated
memory time scale. QUAPI now defines an object called the reduced
density tensor, which lives on this memory time window and establishes an iteration scheme
in order to extract the time evolution of this object. Within the memory time window,
all correlations are included exactly over the finite memory time
$\tau_{\rm mem} = K \Delta t$ and can safely be neglected for times beyond $\tau_{\rm mem}$.
Then, the memory parameter $K$ has to be increased, until convergence is found.
Typical values, for which convergence can be achieved for our problem,
are $K\le 12$ and a reasonable choice is $\Delta t \sim (0.1 - 0.2)/\Delta_0$.
The two strategies to achieve convergence, namely decreasing $\Delta t$ and at
the same time increasing the considered memory time $\tau_{\rm mem} = K \Delta t$, are
clearly increasing both the needed $K$ which results in severe demands considering that the
needed computer power grows exponentially with $K$. Nevertheless convergent results can
be obtained in a wide range of parameters.

\begin{figure}[t]
\epsfig{file=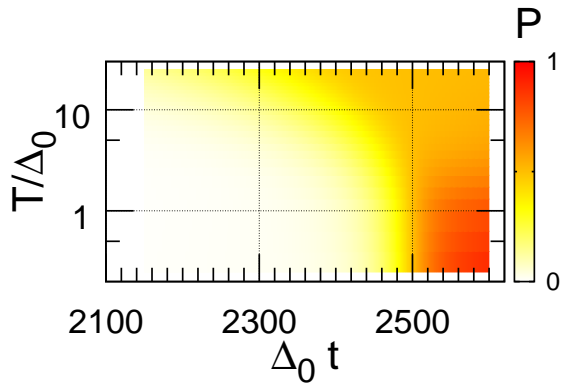,width=8cm}
\vspace*{-1em}\caption{\label{figH2} The occupation probability $P(t)$ of the diabatic
state $|\uparrow\rangle$ versus time and temperature is shown for a slow sweep
velocity $v=0.02 \Delta_0^2$, a weak system bath coupling $\alpha=0.0016$ and a large
cut-off frequency $\omega_c=10 \Delta_0$.
The time is restricted around the crossover which was at $t_{\rm cross}=2500/\Delta_0$.}
\end{figure}

\begin{figure}[t]
\epsfig{file=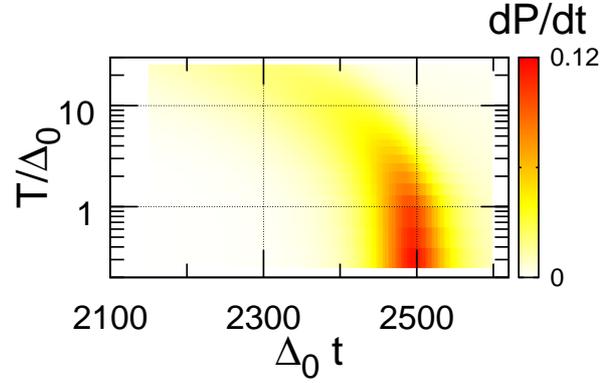,width=8cm}
\vspace*{-1em}\caption{\label{figH3} The time derivative $dP/dt$ corresponding
to Fig.\ \ref{figH2}.}
\end{figure}

Fig.\ \ref{figH2} shows the population $P(t)$ vs.\ time for different temperatures for a fixed sweep
velocity $v=0.02 \Delta_0^2$ in the weak coupling regime $\alpha=0.0016$. We start in the
infinite past with $P(t=-\infty)=0$ as the state $|\downarrow\rangle$ is fully populated, see
Fig.\ \ref{figH1}. The Landau-Zener sweep reaches the minimal gap at $t=2500/\Delta_0$. Approaching this
point, $P(t)$ starts to increase. For larger temperatures, this increase is less pronounced than
for lower temperatures. This is shown more explicitly in Fig.\ \ref{figH3}, where the corresponding
derivative $dP(t)/dt$ is shown. Note that this quantity may be viewed as tunneling rate for the
dissipative Landau-Zener transition. It is naturally more pronounced at low temperatures.
 Increasing the speed of the Landau-Zener sweep, the
population $P(t)$ of the $|\downarrow\rangle$ state shows some transient oscillatory
dynamics before a stationary value is reached (results not shown). We would like
to point out that the iterative QUAPI approach by construction allows the access to
the full time-dependent Landau-Zener transition. In the following, we focus on the
stationary populations at asymptotic times without discussing how the stationary state
is reached in all studied parameter configurations.

\section{Landau-Zener and excitation survival probability}
In this section, we turn to the asymptotic populations and study the Landau-Zener and the excitation
survival probabilities.
\subsection{Landau-Zener probability at weak coupling}

\begin{figure}[t]
\epsfig{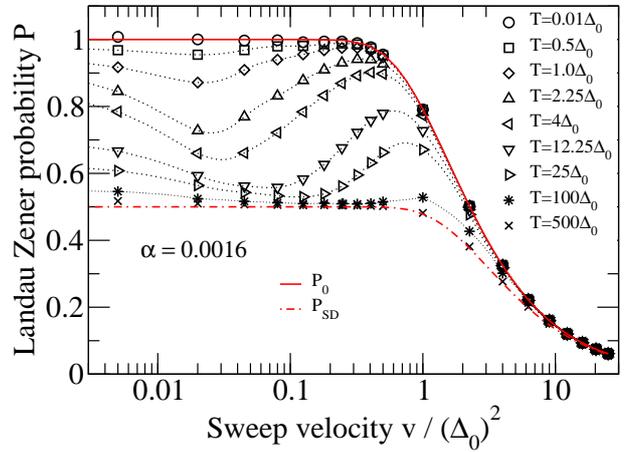}
\vspace*{-1em}\caption{\label{figH5} The Landau-Zener probability $P$ for
various temperatures $T$ is shown for a weak system-bath coupling
$\alpha=0.0016$. The dotted lines are guides to the eye. The solid line marks
the coherent Landau-Zener probability $P_0$, while the dot-dashed line indicates the
high-temperature limit $P_{SD}$, see text.}
\end{figure}

Figure \ref{figH5} shows the Landau-Zener probability $P$ versus sweep velocity
$v$ for different temperatures and for weak coupling, $\alpha=0.0016$. Clearly
the regime with large velocities, $v\gg \Delta_0^{2}$ is distinguishable from a
regime with small velocities (adiabatic regime) $v\lesssim\Delta_0^{2}$.
At low temperatures one expects no sizable influence of the bath
\cite{LZKaI1984,LZKaII1984,LZAo1991,LZKa1998} which should vanish totally at
$T=0$ \cite{LZWu2006,LZSa2007}. This is confirmed by our numerical results.
For small $v$ and low temperatures, $T\lesssim\Delta_0$, we find $P\sim 1$, and
thus unmodified compared to the pure quantum mechanical Landau-Zener result
$P_0$ (solid line).
For increasing velocity, the Landau-Zener probability decreases rapidly and there is
hardly any temperature effect in the considered temperature range, see Fig.\
\ref{figH5}.
This observation agrees with results by Kayanuma and Nakayama who determined the
Landau-Zener probability in the limit of high temperatures,
$P_{SD}=\halb(1-\exp(-\pi\Delta^2/v))$ \cite{LZKa1998,LZKaI1984,LZKaII1984}
(dot-dashed line),
assuming dominance of phase decoherence over dissipation.
For large velocities, $P_{SD}$ decreases as the pure Landau-Zener probability,
$P_{SD}\sim P_{LZ}\sim \halb\pi\Delta^2/v$, and accordingly no sizable
temperature effect is expected.

In the experimentally most relevant parameter range of intermediate to high
temperatures, $T>\Delta_0$ and small sweep velocities, $v<\Delta_0^{2}$, we find
(as reported before~\cite{LZNa2009}) a nontrivial and unexpected behavior of
$P$.
Besides an overall decrease of $P$ with increasing temperature, we find (at
fixed $T$) for decreasing velocity first a maximum of $P$ at $v_{\rm
max}\lesssim \Delta_0^{2}$, then a minium at $v_{\rm min}$ and finally again an
increase.
For decreasing temperatures, $v_{\rm min}$ decreases, and $P(v_{\rm min})$
increases. For high temperatures $T\ge 500\Delta_0$, 
our data follow nicely the predictions by Kayanuma and 
Nakayama~\cite{LZKa1998,LZKaI1984,LZKaII1984}.

This nonmonotonic behavior cannot be described in terms of perturbative
approaches.
Ao and Rammer derived temperature-dependent corrections to the Landau-Zener
probability for low temperatures \cite{LZAo1991}. They report an onset
temperature $T_o\propto 1/v$,
above which temperature affects $P$. Thus, at larger velocities, the decrease of
$P$ due to increasing temperature starts at higher temperatures. This is in line
with our findings of the maximum in $P(v)$, but it does not account for the
minimum and the subsequent increase of $P$ for smaller $v$. In the limit of high
temperatures, $P_{SD}\rightarrow 1/2$ for $v<\Delta_0^2$. Thus, $P_{SD}$
captures the decrease of $P$ in Fig.~\ref{figH5} with increasing temperature,
but it does not account for the nonmonotonic behavior for decreasing $v$.

\begin{figure}[t]
\epsfig{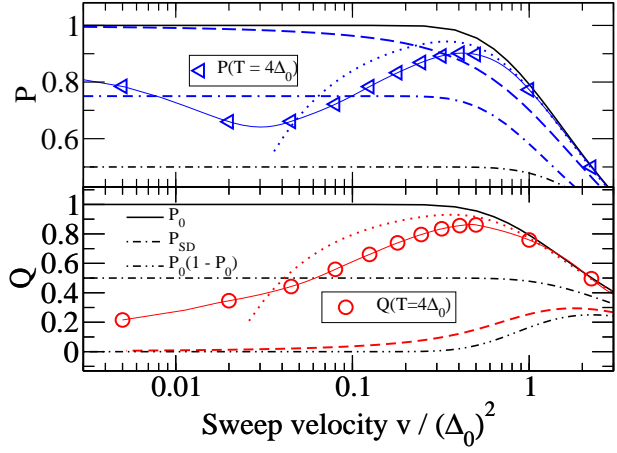}
\vspace*{-1em}\caption{\label{figH7} Comparison
of the Landau-Zener ($P$, top panel, triangles) and the excitation survival
($Q$, bottom panel, circles) probability with perturbative approaches
for $T=4 \Delta_0$ for a weak system-bath coupling $\alpha=0.0016$.
The thick solid line in both panels marks the standard Landau-Zener probability $P_0$,
while the dot-dashed line shows $P_{\rm SD}$.
The dotted lines show the behavior
according to Ao and Rammer \cite{LZAo1991}, the dashed
lines according to Kayanuma and Nakayama \cite{LZKa1998}.
Moreover, in the top panel,
the dash-dash-dotted line marks the result of Pokrovsky and Sun \cite{LZPo2007} (which
only give a formula for the Landau-Zener probability $P$).
In the lower panel, the dot-dot-dashed line indicates $P_0(1-P_0)$. }
\end{figure}

In Fig.~\ref{figH7}, we compare our data (triangles for $P$ (top) and circles for $Q$ (bottom))
for $T=4 \Delta_0$ and $\alpha=0.0016$ with the result of Ao and Rammer
(dotted lines). In fact, the latter describes
qualitatively the maximum both for $P$ and $Q$, but fails quantitatively.
Similarly, we were not able to match either Eq.\ (54) of Ref.\ \cite{LZKa1998}
(dashed line in Fig.~\ref{figH7}) nor Eq.\ (40) of Ref.\ \cite{LZPo2007}
(dash-dash-dotted line in Fig.~\ref{figH7}) with our exact data, see Fig.~\ref{figH7}.
Both describe a
reduction of the Landau-Zener probability with increasing temperature but
neither the maximum nor the subsequent minimum
in the behavior versus $v$ is predicted correctly.

\subsection{Physical picture}
\label{physpic}
The observed behavior can be understood within a simple physical picture
realizing that the bath induces relaxation as main effect, whose time scale can compete
with the Landau-Zener sweep velocity.
Since initially the system is in the ground-state, only absorption can occur, if
an excitation with energy $\Delta_t=\sqrt{\Delta_0^2+(vt)^2}$ exists in the
bath spectrum and is thermally populated. Since $\Delta_t$ is (slowly)
changing with time, relaxation can only occur during a time window $|t|\le\halb
t_r$ with the resonance time
\be \label{restime}
t_r \,=\, \frac{2}{v}\sqrt{\Delta_c^2-\Delta_0^2} \, ,
\ee
in which the energy splitting fulfills the condition
$\Delta_{t}\le\Delta_c=\mbox{min}\{T,\omega_c\}$ \cite{note1} as illustrated in
Fig.~\ref{figH1}.
In order for relaxation processes to contribute, the (so far unknown) relaxation
time $\tau_r$ must be shorter than $t_r$.

\begin{figure}[t]
\epsfig{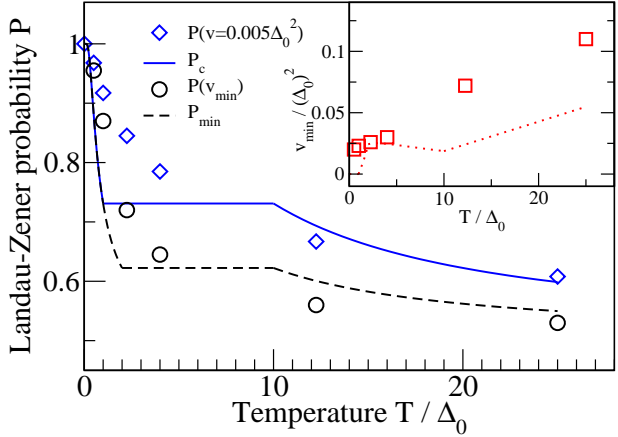}
\vspace*{-1em}\caption{\label{figH15}Main: Landau-Zener probability $P(v=0.005\Delta_0^2)$
at the smallest sweep velocity (blue diamonds), $P_c$ of Eq.~(\ref{pc}) (blue solid line),
Landau-Zener probability $P(v_{\rm min})$ at the sweep velocity of the minimum (black circles),
and $P_{\rm min}$ of Eq.~(\ref{pvmin}) (back solid line) versus temperature. Inset:
The minimum sweep velocity (red squares) extracted from the data of
Fig.~\ref{figH5} and $v_{\rm min}$ according to Eq.~(\ref{vmin2}) (red dotted line).
The coupling strength is $\alpha=0.0016$.}
\end{figure}

For large sweep velocities, $t_r\ll\tau_r$, relaxation is negligible and no
influence of the bath is found as expected. In the opposite limit,
$t_r\gg\tau_r$, relaxation will dominate and the two levels will at any time
adjust their occupation to the momentary $\Delta_t$ and $T$. Once
$\Delta_t\ge\Delta_c$, relaxation stops since no spectral weight of the bath
modes is available and the corresponding
``critical'' Landau-Zener probability can be estimated as
\be \label{pc}
P_c \,=\, \halb \left[ 1+\tanh\left(\frac{\Delta_c}{2 T} \right) \right] .
\ee
For small but finite $v$, equilibration is retarded. The two levels need
the finite relaxation time to adjust their occupation to the momentary $\Delta_t$
and $T$. In this time, however, $\Delta_t$ might
exceed $\Delta_c$ and then relaxation is not possible anymore. Thus equilibrium is
reached for an energy splitting in the past $\Delta_{t'}<\Delta_{t_c}$ with $t'<t_c$
and $t_c$ the time when $\Delta_{t_c}=\Delta_c$.
Accordingly, $P$ increases with decreasing $v$ since $\Delta_t$ changes slower and
$P(v\rightarrow0)\le
P_c$, as observed in Fig.~\ref{figH5}. In Fig.~\ref{figH15} (main), we plot the
Landau-Zener probability $P(v=0.005\Delta_0^2)$ (blue diamonds) for the smallest investigated sweep velocity
and compare it with $P_c$ of Eq.~(\ref{pc}) (blue full line).

Relaxation will maximally suppress the Landau-Zener transition when both time scales
coincide, leading to a minimum of $P$ at $v_{\rm min}$ given by the condition
\be\label{vmin} t_r(v_{\rm min})=\tau_r(T,\alpha,\omega_c).
\ee
Within the resonance time window, $|t|\le\halb t_r$, only a single phonon absorption is likely.
We thus can assume equilibration associated to a time-averaged energy splitting
$\overline{\Delta}_r=(2/t_r)\int_0^{t_r/2}dt\Delta_t\simeq\mbox{max}\{\Delta_c/2
,\Delta_0\}$
and a resulting Landau-Zener probability
\be\label{pvmin} P_{\rm min} \,=\, \halb
\left[1+\tanh\left(\frac{\overline{\Delta}_r}{2T}\right)\right] = P(v_{\rm min}).
\ee
Subsequently, there is a maximum for a sweep velocity between $v_{\rm
min}<v_{\rm max}<\Delta^2_0$.
Fig. \ref{figH15} (main) shows the QUAPI data $P(v_{\rm min})$ (black circles) vs. $T$
together with $P_{\rm min}$ given in Eq.~(\ref{pvmin}) (black dashed
line).

For a fixed time and for weak coupling, we can estimate the decay rate out of
the ground state using Golden Rule, $\tau^{-1}(t)=\pi\alpha(\Delta_0^2/\Delta_t)
\exp(-\Delta_t/\omega_c) n(\Delta_t)$ with the Bose factor
$n(\Delta_t)=[\exp(\Delta_t/T)-1]^{-1}$.
For the time-dependent Landau-Zener problem at slow sweep velocities, we may
assume that the bath sees a time-averaged two-level system and thus estimate
the relaxation rate $\tau_r^{-1}$ by using the time-averaged energy splitting
$\overline{\Delta}_r$, i.e.,
\be\label{rate} \tau_r^{-1} \,\simeq\, \pi\alpha
\frac{\Delta^2_0}{\overline{\Delta}_r}\, \exp(-\overline{\Delta}_r/\omega_c)
n(\overline{\Delta}_r)\, .
\ee
For increasing temperature, relaxation becomes faster, and, accordingly, the
condition for $v_{\rm min}$, Eq.~(\ref{vmin}), leading to
\be\label{vmin2} v_{\rm min} \,=\, 2\tau_r^{-1} \sqrt{\Delta_c^2-\Delta_0^2}
\ee
is fulfilled for larger velocities. Qualitatively this picture describes the
temperature dependence of $v_{\rm min}$ observed in the inset of
Fig.~\ref{figH15}. We plot $v_{\rm min}$ (red squares) taken from the data of
Fig.~\ref{figH5} and $v_{\rm min}$ according to Eq.~(\ref{vmin2}).
In detail, given the section-wise definition of the parameters $\Delta_c$ and
$\overline{\Delta}_r$, the agreement in Fig.~\ref{figH15} is rather satisfactory,
keeping in mind that there are no adjustable parameter involved.

\subsection{Excitation survival probability at weak coupling}

\begin{figure}[t]
\epsfig{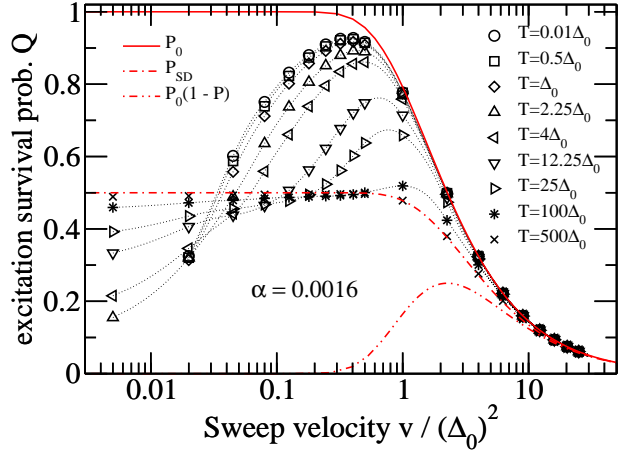}
\vspace*{-1em}\caption{\label{figH6} The excitation survival probability $Q$ for
various temperatures $T$ is shown for a weak system-bath coupling
$\alpha=0.0016$. The dotted lines are guides to the eye. The solid line marks
the coherent excitation survival probability $Q_0=P_0$ which is identical to the
Landau-Zener probability. The dot-dashed line indicates the high-temperature
limit $P_{SD}$ and the dot-dot-dashed line indicates the strong coupling limit
$Q_{\rm sc}=P_0(1-P_0)$.}
\end{figure}

Figure~\ref{figH6} shows the excitation survival probability $Q$
versus sweep velocity $v$ for different temperatures and
for weak coupling, $\alpha=0.0016$. As for
the Landau-Zener probability, the regime with large velocities, $v\gg
\Delta_0^{2}$ is distinguishable from a regime with small velocities (adiabatic
regime) $v\lesssim\Delta_0^{2}$.
For large sweep velocities, the excitation survival probability decreases
rapidly. There is no sizable difference to the undamped case
for all the considered temperatures, as expected in the
high temperature limit, $Q_{SD}(v\gg \Delta_0^{2})$ as well as in the low
temperature limit for weak coupling $Q_0(v\gg \Delta_0^{2})$ and strong coupling
$Q_{\rm sc}(v\gg \Delta_0^{2})$ since
\bee Q_{\rm sc}(v\gg \Delta_0^{2}) &\simeq& Q_0(v\gg \Delta_0^{2}) \,\simeq\,
Q_{SD}(v\gg \Delta_0^{2}) \nonumber\\
&\simeq& P_{0}(v\gg \Delta_0^{2})\,\simeq\, \halb\pi\frac{\Delta^2}{v}
\eee

At low temperatures, $T\lesssim\Delta_0$, no sizable influence of the bath is
found as the data in Fig.~\ref{figH6} for $T=0.01\Delta_0$ almost coincide with
the data for $T=0.5\Delta_0$ (and even more so for data at $T=0.05\Delta_0$ and
$T=0.1\Delta_0$ not shown in the figure). However, for all temperatures we find
the excitation survival probability to peak for sweep velocities between $(0.1 -
1)\Delta_0^2$ and a strong decrease for lower sweep velocities as predicted by
\cite{LZAo1991,LZKa1998}. For small sweep velocities relaxation of the excited
state takes place and thus reduces the excitation survival probability as soon
as the relaxation time becomes comparable or shorter than the resonance time.
Due to spontaneous emission relaxation is present at all temperatures.
Furthermore we would expect in this picture that only at temperatures
$T\ge\Delta_0$ the relaxation time becomes shorter due to induced emission by
excited phonons and thus only for $T\ge\Delta_0$ a temperature effect should be
visible in the excitation survival probability. This is confirmed
by the results in Fig.~\ref{figH6}. Consistent with our simple physical picture
introduced in the last section is that the sweep
velocity of the sharp reduction of the excitation survival probability coincides
roughly with the minimum in Fig.~\ref{figH5} for $T>\Delta_0$. It, as well,
shifts to larger $v$ for increasing temperatures.

\begin{figure}[t]
\epsfig{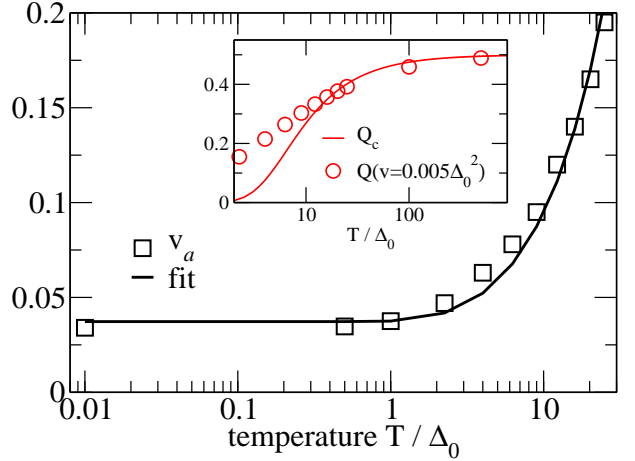}
\caption{\label{figHB2}Main: Sweep velocity $v_a=v[Q(T)=\halb\Delta_0]$
(with temperature fixed) for which the resonance time (Eq.~(\ref{restime}))
coincides with the relaxation time (up to a prefactor $f_a$) vs.\ temperature. Inset:
Critical excitation survival probability $Q_c$, Eq.\ (\ref{qc}) (solid line),
and $Q(v=0.005\Delta_0^2)$ extracted from
Fig.~\ref{figH6}, versus temperature for $\alpha=0.0016$.
}
\end{figure}

Taking the sweep velocity $v_a=v[Q(T)=\halb\Delta_0]$ (with temperature fixed) as
a measure for the {\it sweep velocity of the sharp reduction}, we expect $v_a$ to
be the sweep velocity for which the resonance time (Eq.~(\ref{restime}))
coincides with the relaxation time, up to a prefactor $f_a$ of order one which
is due to the arbitrary definition of $v_a$, i.e.,
\be \label{form2}
f_a t_r(v_a) \,=\, \tau_{d,r} \, .
\ee
The inverse decay time of the excited state is now given as
\be\label{downrate} \tau_{d,r}^{-1} \,\simeq\, \pi\alpha
\frac{\Delta^2_0}{\overline{\Delta}_r}\, \exp(-\overline{\Delta}_r/\omega_c)
\left[ 1 + n(\overline{\Delta}_r)\right] \, .
\ee
Since thermal occupation is no limiting factor for the decay of the excited state,
we have that $\Delta_c=\omega_c$ and thus $\overline{\Delta}_r=\omega_c/2$.
Fig.~\ref{figHB2}
plots $v_a$ (black squares) versus temperature which nicely shows a temperature
behavior $\sim [1 + n(\overline{\Delta}_r)]$. The black line marks the result
of the best fit of Eq.~(\ref{form2}) with the
fitting parameter $f_a\simeq 3.04$, confirming again the good
agreement between our numerical data and our simple physical picture.

In the high temperature limit, $Q\rightarrow P_{SD}$~\cite{LZAo1991,LZKa1998}.
In our data, we observe that the peak in $Q$ decreases with increasing
temperature and $Q$ at smallest investigated sweep velocity increases with
increasing temperature. Both features are in agreement with the expectation in
the high temperature limit. For the smallest investigated sweep velocity,
the relaxation time should exceed the resonance time by far and thus we expect within our simple
picture that the system fully relaxes towards the momentary equilibrium until
the energy splitting exceeds $\Delta_c=\omega_c$. This yields the condition
\be \label{qc}
Q_c \,=\, \halb \left[ 1-\tanh\left(\frac{\Delta_c}{2 T} \right) \right] .
\ee
The inset of
Fig.~\ref{figHB2} compares $Q_c$ with $Q(v=0.005\Delta_0^2)$ extracted from
Fig.~\ref{figH6} and again shows satisfactory agreement.

The excitation survival probability is numerically much harder to determine
than the
Landau-Zener probability. For example, at the sweep velocities
$v=0.005\Delta_0^2$ our data did not converge for temperatures below
$T=4\Delta_0$ and hence is not shown. According to our physical picture,
the main effects of the bath occur during the resonance time window. Outside of it,
 only multi-phonon processes can occur, which are
strongly suppressed due to the weak coupling. This directly affects our numerical
scheme which is thus not sensitive to the very long times needed to perform a Landau-Zener
experiment (from $t=-\infty$ to $t=+\infty$). In contrast,
for the Landau-Zener probability
all relaxation effects outside the resonance time window are additionally suppressed
by thermal occupation numbers of the phonons. This effect does not influence the
excitation survival probability due to spontaneous decay. These multi-phonon
processes outside the resonance time window are, however, processes which need long
memory in the bath and may thus be critical for convergence of our numerics. We should
emphasize that our simple physical picture even explains our stronger
numerical efforts to obtain
converged results for the excitation survival probability.

In summary, at weak coupling our simple physical picture explains qualitatively fully and quantitatively satisfactorily both the Landau-Zener as well as the excitation survival probability. Both are unmodified by the dissipative effects of the environment for large velocities, $v\gg\Delta_0^{2}$. For slow velocities , $v\lesssim\Delta_0^{2}$, both are strongly influenced due to relaxation where a striking difference emerges since the excited state can relax by spontaneous emission whereas the ground state can only absorb a phonon when one is thermally occupied. Thus the Landau-Zener probability at vanishing temperature is unmodified by the presence of the bath whereas the excitation survival probability is strongly altered, i.e. decreases rapidly for sweep velocities smaller than $v_a$ determined by the competition of driving and relaxation.

\subsection{Medium to strong couplings}

\begin{figure}[t]
\epsfig{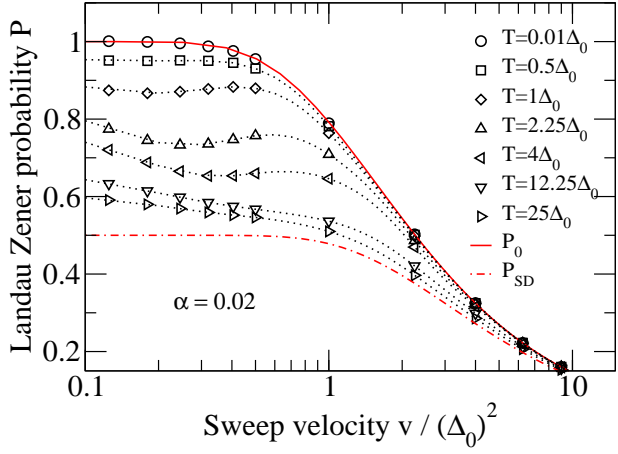}
\vspace*{-1em}\caption{\label{figH8} The Landau-Zener probability $P$ for
various temperatures $T$ is shown for an intermediate value of the
 system-bath coupling $\alpha=0.02$. The
dotted lines are guides to the eye. The solid line marks the coherent
Landau-Zener probability $P_0$, while the dot-dashed line indicates the
high-temperature limit $P_{SD}$.}
\end{figure}

\begin{figure}[t]
\epsfig{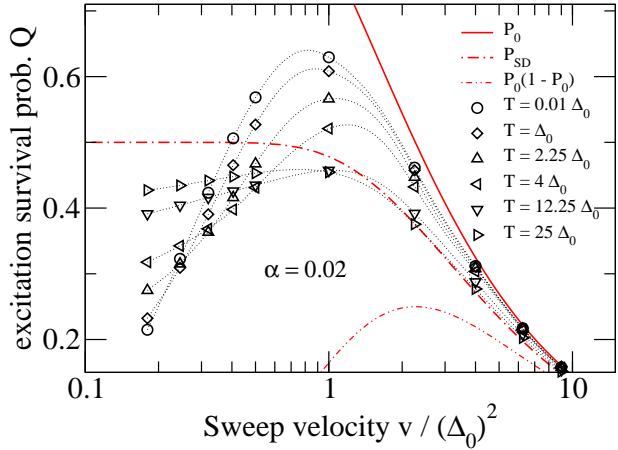}
\vspace*{-1em}\caption{\label{figH9} The excitation survival probability $Q$
corresponding to the case shown in Fig.\ \ref{figH8}. Here,
the dot-dot-dashed line marks the strong coupling limit $Q_{\rm
sc}=P_0(1-P_0)$.}
\end{figure}

\begin{figure}[t]
\epsfig{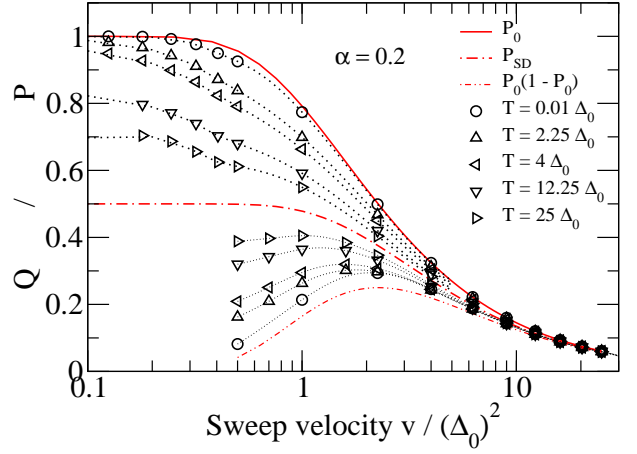}
\vspace*{-1em}\caption{\label{figH10} The
Landau-Zener ($P$, all data above the red dash-dotted line) and the excitation
survival ($Q$, all data below the red dash-dotted line) probability for various
temperatures $T$ are shown for strong system-bath coupling $\alpha=0.2$. The
dotted lines are guides to the eye.
The solid line marks the coherent excitation probability which is identical to the coherent
Landau-Zener probability $P_0$.}
\end{figure}

Surprisingly, our simple picture still holds qualitatively for stronger damping,
when the Golden Rule is not expected to hold.
Increasing $\alpha$ enhances relaxation and $\tau_r$ and $\tau_{d,r}$ decrease.
Thus, the sweep velocity of the minimum in the Landau-Zener probability as well
as $v_a$ in the excitation survival probability increase for larger coupling
strengths $\alpha$ (at fixed temperature).
This is confirmed by Fig.~\ref{figH8} where $P$ is shown for
the same temperatures as in Fig.~\ref{figH5}, but for a larger value of
$\alpha=0.02$. Fig.~\ref{figH9} shows the corresponding excitation
survival probability for $\alpha=0.02$.
The minimum of $P$ and $v_a$ for $Q$ are still observable for $\alpha=0.02$ for
temperatures $\Delta_0\lesssim T\lesssim 4\Delta_0$.
At higher temperatures, only a shoulder remains for $P$ in Fig.~\ref{figH8} and
$Q$ in Fig.~\ref{figH9} does not exceed $1/2$ anymore.
Results for $P$ and $Q$ for even stronger coupling $\alpha=0.2$ are shown in
Fig.~\ref{figH10}. For $\alpha=0.2$, the local extrema in $P$ disappear, but a monotonic growth of
$P$ with decreasing sweep velocity is still in line with our simple picture. For
coupling strengths $\alpha\ge1/\sqrt{2}$ no bath influence is expected
anymore~\cite{LZAo1991} for the Landau-Zener probability, consistent with our
data. Therefore, we focus on $\alpha\le0.2$. At the same time the excitation
survival probability is expected to follow $Q_{\rm sc}=P_0(1-P_0)$ at strong
coupling and low temperatures~\cite{LZKa1998}. As seen in Fig.~\ref{figH10}, $Q$
indeed approaches $Q_{\rm sc}$ with decreasing temperature for $\alpha=0.2$.

This striking behavior of the excitation survival probability at strong
coupling opens a road to determine the coupling strengths in experimental
systems. Under weak coupling conditions, the minimum in the Landau-Zener probability
allows to obtain $\alpha$. For strong coupling, the minimum vanishes but the
excitation survival probability still shows a clear peak. Especially the
temperature dependence of the peak clearly separates strong from weak coupling.
For strong coupling, the peak height increases with temperature while for
weak coupling, the peak height decreases with
increasing temperature.

\subsection{Dependence of $P$ and $Q$ on the system-bath coupling}

\begin{figure}[t]
\epsfig{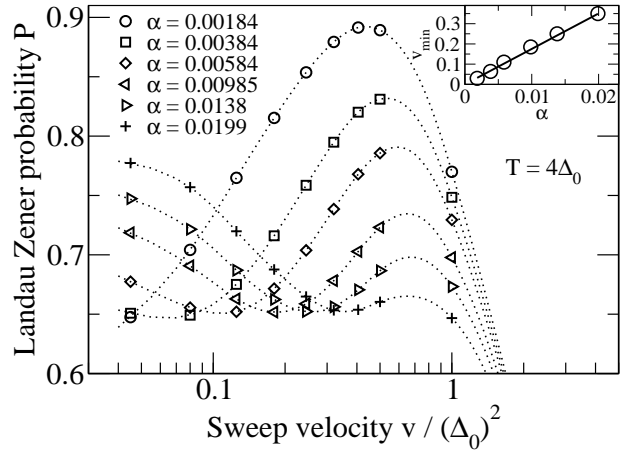}
\vspace*{-1em}\caption{\label{figH11} Landau-Zener probability $P$ for different $\alpha$
for $T=4\Delta_0$. Inset: Linear dependence of
$v_{\rm min}$ on $\alpha$, see text.}
\end{figure}

Next, we investigate the dependence of $P$ and $Q$ on the system-bath coupling
in the regime of weak coupling.
Fig.\ \ref{figH11} shows the Landau-Zener probability for different $\alpha$ at
a fixed temperature $T=4 \Delta_0$. For increasing coupling, $v_{\rm min}$
shifts to larger velocities. In fact, $v_{\rm min}$ depends linearly on
$\alpha$, see inset of Fig.\ \ref{figH11}.
The linear dependence is also predicted by our model, i.e., $v_{\rm
min}=15.8\alpha\Delta_0^2$,
in very good agreement with the fit $v_{\rm min}=17.54\alpha\Delta_0^2$.
The decreasing maximum $P(v_{\rm max})$ results from the shifting minimum.
Another remarkable fact is that the Landau-Zener probability $P(v_{\rm min})$ at
the minimum velocity is independent of $\alpha$, as predicted by our
physical picture, see Eq.~(\ref{pvmin}). We estimate the averaged splitting
$\overline{\Delta}_r=T/2=2\Delta_0$, in fair agreement with
$\overline{\Delta}_r=2.476\Delta_0$, obtained with $P(v_{\rm min})=\halb
[1+\tanh(\overline{\Delta}_r/2T)]$ from the data in Fig.\ \ref{figH11}. This
agreement strongly supports our physical picture that relaxation dominates in
the intermediate temperature range for small sweep velocities.

\begin{figure}[t]
\epsfig{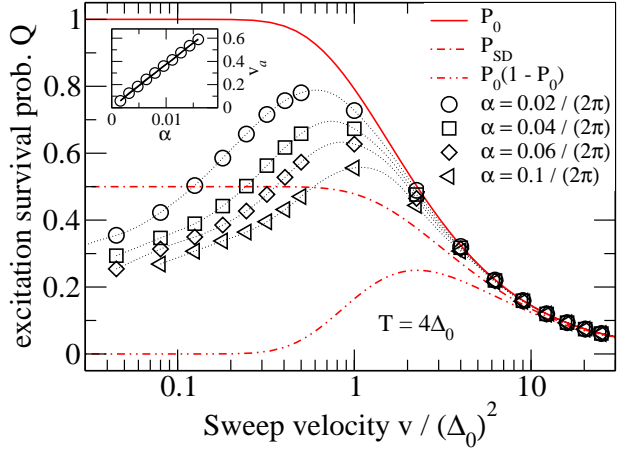}
\vspace*{-1em}\caption{\label{figH12} Excitation survival probability $Q$ for different $\alpha$
for $T=4\Delta_0$. Inset: Linear dependence of $v_a$ on $\alpha$, see text.}
\end{figure}

The excitation survival probability $Q$ is shown in Fig.~\ref{figH12} for
different $\alpha$ at a fixed temperature $T=4 \Delta_0$. For increasing
coupling, $v_a$ shifts to larger velocities. As before $v_{\rm min}$ for the 
Landau-Zener probability, now $v_a$ for the excitation survival probability
 depends linearly on $\alpha$, see inset of Fig.~\ref{figH12}.
Again, the linear dependence is predicted by our model, i.e.,
$v_a=32.3\alpha\Delta_0^2$ (taking the previously determined factor $f_a=3.04$
into account),
in very good agreement with the fit $v_{\rm min}=37.4\alpha\Delta_0^2$.
The decreasing peak height in $Q$ results from the shifting $v_a$.

\subsection{Dependence on cut-off frequency}

\begin{figure}[t]
\epsfig{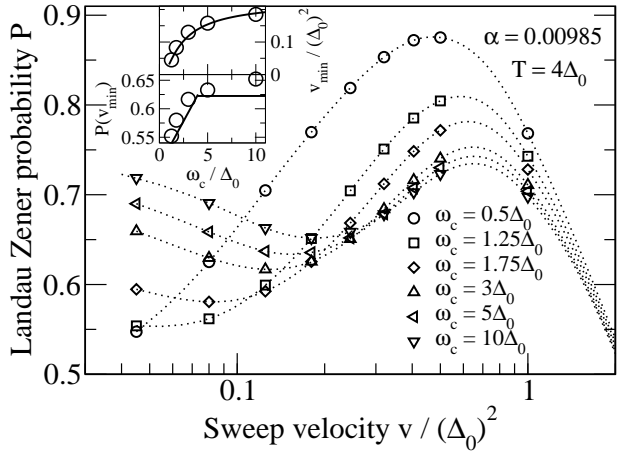}
\vspace*{-1em}\caption{\label{figH13} Landau-Zener probability $P$ for various cut-off
frequencies $\omega_c$ for $T=4\Delta_0$ and $\alpha=0.00985$. Insets: $P(v_{\rm
min})$ (bottom) and $v_{\rm min}$ (top) vs $\omega_c$; solid lines:
model predictions, see text.}
\end{figure}

\begin{figure}[t]
\epsfig{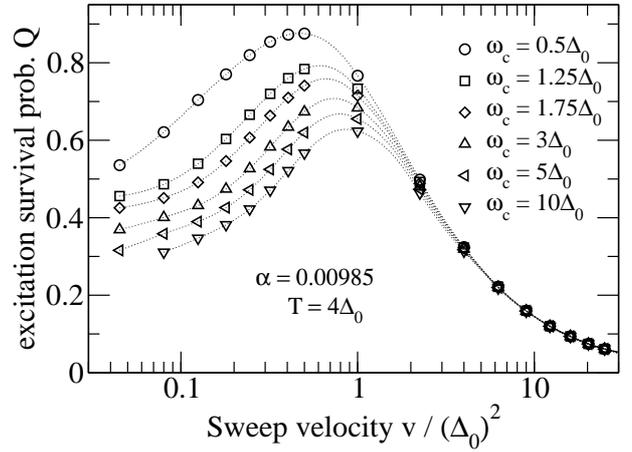}
\vspace*{-1em}\caption{\label{figH14} Excitation survival probability $Q$
corresponding to Fig.\ \ref{figH13}.}
\end{figure}

There is an additional time scale provided by the bath dynamics, which
determines how fast the bath relaxes to its own thermal equilibrium
due to the coupling to the system. It is given by the reorganization energy
\cite{WeissBuch} and depends on the cut-off frequency $\omega_c$
of the bath spectrum. In turn, the relaxation rate
(\ref{rate}) also depends on $\omega_c$ and relaxation is strongly suppressed when
$\Delta_t>\omega_c$.
Fig.\ \ref{figH13} shows $P$ and Fig.\ \ref{figH14} shows $Q$ for different $\omega_c$, 
ranging down to $\omega_c =0.5 \Delta$. Such small values of the cut-off
frequency describe slow bath fluctuations, a situation, for instance,
 typical for the biomolecular exciton dynamics in a protein-solvent
environment \cite{BioTh2008}.
With decreasing cut-off frequencies, the minimum in $P(v)$ as well as $v_a$ in $Q$ 
shift to smaller $v$ as qualitatively expected from Eq.\ (\ref{rate}) and (\ref{downrate}) respectively.
We note that this is rather
surprising since a small $\omega_c$ also induces strong non-Markovian effects.
At the same time, the Landau-Zener probability $P(v_{\rm min})$ decreases.
With decreasing $\omega_c$, the resonance time $t_r$ and the averaged energy
splitting $\overline{\Delta}_r$ also decrease. For cut-off frequencies
$T\le\omega_c$, we expect $v_{\rm min}=0.19\exp(-2\Delta_0/\omega_c)$, in fair
agreement with the fit $v_{\rm min}=0.23\exp(-2\Delta_0/\omega_c)$. This is
shown in the upper inset in Fig.\ \ref{figH13}. The lower inset shows $P(v_{\rm
min})$ versus $\omega_c$. The solid lines are predictions from our model and are
in fair agreement with data. A similar analysis for $v_a$ was not possible since for 
small $\omega_c$ the excitation survival probability $Q(v)$ did not 
sharply fall below $\halb$ and thus $v_a$ could not be determined unambiguously.

\section{Summary}

We have investigated the dissipative Landau-Zener problem by means of the numerically
exact quasiadiabatic propagator path-integral \cite{QUAPI1,QUAPI2,QUAPI3}
approach for an Ohmic bath.
Thereby we discussed the Landau-Zener probability (to end up in the ground state
when starting in the ground state) as well as the excitation survival probability (to end up in the excited state when starting in the excited state).
In the limits of large and small sweep velocities and low temperatures, our results coincide with
analytical predictions \cite{LZAo1991,LZKa1998,LZWu2006,LZZu2008}.
In the intermediate regime, when the
sweep velocities are comparable to the minimal Landau-Zener gap and intermediate temperatures, we have identified novel
non-monotonic dependencies of the Landau-Zener probabilities
on the sweep velocity, temperature, system-bath coupling strength
and cut-off frequency. This parameter range is clearly not accessible
by perturbative means.
The observed behavior can be understood in rather simple physical terms as a nontrivial
competition between relaxation and Landau-Zener driving.
The main difference between the Landau-Zener and the excitation survival probability results from the simple fact that the excited state can always decay via spontaneous emission while the ground state needs a thermally excited phonon in order to become excited. Thus even at vanishing temperature the excitation survival probability decreases strongly for small enough driving speed whereas the Landau-Zener probability only for high temperatures.
As nowadays advanced experimental set-ups allow for a rather
comprehensive control of the parameters, this
novel feature should be accessible by available experimental techniques.

We thank V. Peano and S. Ludwig for discussions and acknowledge support by the
Excellence Initiative of the German Federal and State Governments.


%




\begin{thebibliography}{99}
\bibitem{note1} For $\omega_c, T<\Delta_0$, our definition of the resonance time
is inept and setting $\Delta_c=\Delta_0$ is more adequate.
\end{thebibliography}
\end{document}